\documentclass[aps,showpacs,amsfonts,amsmath, amssymb,twocolumn,floatfix,superscriptaddress]{revtex4-1}
\usepackage[final]{graphicx}
\usepackage{color}
\usepackage{amssymb,amsmath}

\newcommand{\vct}[1]{\mathbf{#1}}

\newcommand{\be}{\begin{equation}}
\newcommand{\ee}{\end{equation}}
\allowdisplaybreaks

\DeclareSymbolFont{bbgreek}{U}{bbold}{m}{n}
\DeclareMathSymbol{\bbmu}{\mathbb}{bbgreek}{'26}
\DeclareMathSymbol{\bbeps}{\mathbb}{bbgreek}{'17}
\begin{document}

\title{A Gaussian theory for fluctuations in simple liquids}

\date{\today}

\author{Matthias Kr\"uger}
\affiliation{4th Institute for Theoretical Physics, Universit\"at Stuttgart, Germany and Max Planck Institute for Intelligent Systems, 70569 Stuttgart, Germany
}
\author{David S. Dean}
\affiliation{Univ. Bordeaux and CNRS, Laboratoire Ondes et Matière d'Aquitaine (LOMA), UMR 5798, F-33400 Talence, France}

\begin{abstract} 
Assuming an effective quadratic Hamiltonian, we derive an approximate, linear stochastic equation of motion for the density-fluctuations in liquids, composed of overdamped Brownian particles. From this approach, time dependent two point correlation functions (such as the intermediate scattering function) are derived. We show that this correlation function is exact at short times, for any interaction and, in particular, for arbitrary external potentials so that it applies to confined systems. Furthermore, we discuss the relation of this approach to previous ones, such as dynamical density functional theory as well as the formally exact treatment. This approach,  inspired by the well known Landau-Ginzburg Hamiltonians, and the corresponding ``Model~B'' equation of motion, may be seen as its microscopic version, containing information about the details on the particle level.
\end{abstract}

\pacs{
05.40.-a, 
61.20.Gy, 
61.20.Lc, 
82.70.Dd 
}
\bibliographystyle{plain}

\maketitle

\section{Introduction}
The statics and dynamics of simple liquids is of great importance both in fundamental research \cite{HansenMcDonald, bob_review,Goetze}, but also in industry, technology and biology. The statics have been investigated for many years and are well understood, for instance via the framework of classical density functional theory \cite{HansenMcDonald,bob_review,roth_review}.

Studies of the dynamical properties of  fluids, such as the viscosity, have a long history \cite{Maxwell_viscosity}, and the field is still very active \cite{Viscosity_of_Liquids,  ElliottPRL2014ViscFermGasMeasurement}. Linear response theory \cite{Green, Kubo}, connects transport coefficients to time dependent correlation functions measured in thermal equilibrium, the time dependent correlation functions \cite{HansenMcDonald, dhont, Risken} studied here are thus 
of particular importance.. 

Time dependent correlation functions can be computed from various fundamental equations, such as the Liouville \cite{HansenMcDonald} or Fokker Planck equations \cite{Risken, dhont}. Dilute systems have been examined using exact dynamical formulations, for instance via the Boltzmann equation \cite{Kreuzer, Andrade} or using the Fokker Planck equation \cite{Risken, dhont, Pusey}.  In dense systems, approximate dynamical formulations have been used: Here, Mode Coupling Theory (MCT) \cite{Goetze} is useful for the computation of time correlation functions in bulk systems, and has recently also been applied in confinement \cite{Lang10, Lang12,Lang14,Lang14_2}. Classical Density Functional Theory finds static equilibrium quantities \cite{HansenMcDonald,bob_review,roth_review}, while {\it Dynamical} Density Functional Theory (DDFT) \cite{Marconi99, archer, RexLoewenPRL2008} is powerful for describing out of equilibrium situations. In addition to the evolution in time dependent potentials, DDFT has also been used to study driven suspensions with spherical obstacles \cite{penna,rauscher2} or with constrictions \cite{Zimmermann16}, driven liquid crystals \cite{Hartel10}, suspensions under shear \cite{Brader_Kruger_2011, Brader_Kruger_2011_2, Johannes_Joe_2013,AerovKrugerJCP2014,aerov2015} and for microswimmers \cite{Menzel16}. Such research directions have also benefitted from formal improvements within power functional theory \cite{SchmidtBraderJCP_2013_power_func,Brader13}.

Despite these many applications, DDFT provides no immediate access to the time dependent equilibrium correlation functions, however the test particle trick has been used to derive the van Hove function from it \cite{Hopkins10, Archer10,Brader15}. 

On the experimental side, the intermediate scattering function is an important quantity characterizing the dynamics of liquids, e.g.~as regards the glass transition \cite{Megen93}, and can also be measured in confinement \cite{Nygard16}. More generally,   
the dynamics of fluids in confinement have received a lot of recent attention \cite{Petravic2_2006_JCP_viscosity_lin_resp_between_planes, Klapp2008PRL_SimCollFilm, Zhang2004SimPoisFl}, among other reasons due to improved experimental precision on small scales \cite{IgnatovichPRL2006DetectingNanoPart, Isa09, Cheng11, Chevalier}, and in microfluidic devices \cite{Psaltis2006_Nature_microfl_dev, Sajeesh2014_rev_Part_sep_in_Mfl_dev} or blood flow in capillaries \cite{Li_blood_fl_sim, Zhou2005_capillary_blood_susp}.

Previous  approaches that discuss the stochastic dynamics of particle densities, including noise, have been presented in Refs.~\cite{Dean96} and \cite{Archer04b}, see also Ref.~\cite{Donev14}. We will discuss their relation with the approach developed here. 

In this manuscript, we propose a description of fluctuations of fluids near equilibrium by use of a Gaussian field theory, corresponding to an effective  quadratic Hamiltonian for the density fluctuation field and a corresponding Langevin equation. The Hamiltonian is constructed to yield the correct static equilibrium averages. The corresponding Langevin equation is constructed to yield the dynamics of overdamped particles. Within this theory, we derive a closed, approximate expression for the time depend equilibrium correlation function, which agrees with the exact result (found from the Smoluchowski equation) for short times. It is thus expected to describe well the dynamics at not too high densities, and might especially provide insight into  dynamics in confined systems. We also demonstrate the connection between the derived dynamics and the dynamics following from dynamical density functional theory, as well the exact stochastic equation for the density operator.

The manuscript is structured as follows. In Section \ref{sec:H}, we lay out the theoretical framework, starting with the system considered in Section~\ref{sec:S}. We define the physical observables of interest in Sec.~\ref{sec:Ob}. The quadratic effective Hamiltonian is introduced in Section \ref{sec:eH},  and the stochastic equation of motion is introduced in Section~\ref{sec:eom}. The time dependent correlation function from this framework is computed in Section \ref{sec:tdc}, and we demonstrate in Section \ref{sec:ste}, that it is exact for short times. In Section \ref{sec:DDFT}, we show that proposed found equation of motion is in close correspondence with DDFT. Sec.~\ref{sec:SDE} discusses the connection  to the exact equation of motion for the density field of Ref.~\cite{Dean96}. 
We summarize in Section \ref{sec:Su}.  

\section{System, effective Hamiltonian and equation of motion}\label{sec:H}
\begin{table}
\begin{ruledtabular}
\begin{tabular}{|p{2.3cm}|p{5.9cm}|}
\hline
Symbol &Meaning\\
\hline\hline\hline
$\rho(\vct{x})$&Density operator: $\rho(\vct{x})=\sum_i\delta(\vct{x}-\vct{x}_i)$\\
\hline
$\langle\rho(\vct{x})\rangle$&Mean density in equilibrium.\\ 
\hline
$\phi(\vct{x},t)$& Fluctuation of density near about its equilibrium value, $\phi(\vct{x},t)=\rho(\vct{x})-\langle\rho(\vct{x})\rangle$.\\
\hline
$\left\langle \phi(\vct{x},t)\phi(\vct{x}',t')\right\rangle$&Time dependent correlations of density fluctuations in equilibrium, the quantity of interest of this work.\\
\hline
$\langle\rho(\vct{x},t)\rangle^{\rm{neq}}$&Mean density in nonequilibrium state.\\
\hline
$\delta\rho(\vct{x},t)$&Average difference from equilibrium value in a perturbed system, $\delta\rho(\vct{x},t)=\langle\rho(\vct{x},t)\rangle^{\rm{neq}}-\langle\rho(\vct{x})\rangle$.\\
\hline
\end{tabular}
\end{ruledtabular}
\caption{\label{table:1}Observables studied in this manuscript. The lower two rows, i.e., the density in nonequilibrium states, are given for comparison to dynamical density functional theory in Sec.~\ref{sec:DDFT}. }
\end{table}

\subsection{System}\label{sec:S}
We aim to analyze time dependent correlation functions in liquids. For this, we choose a well studied, and also experimentally relevant model system, which is overdamped spherical (Brownian) particles. (The question how well this model system describes aspects of molecular liquids also attracted recent interest \cite{Bollinger15}.)

Regarding the ensemble, using Brownian dynamics directly implies a canonical or grand-canonical description, where the solvent acts as a bath at the given temperature. We will generally have in mind systems for which canonical and grand canonical descriptions are equivalent due to the large (infinite) particle number (such as in the semi-infinite system bound by a planar surface). Extra care has  thus to be taken for closed systems, such as particles confined in a box of finite size (see Ref.~\cite{Heras14} for an analysis of canonical systems in DFT).

The Brownian particles with positions at $\vct{x}_i$ are subject to a potential $\Phi(\{\vct{x}_i\})$, including pairwise interactions (later denoted by $V$) as well as an external potential (later denoted $U$). The thermal energy  scale is denoted by $k_BT\equiv\beta^{-1}$, with Boltzmann constant $k_B$ and the (solvent imposed) temperature $T$. The bare diffusivity (the diffusivity in the absence of interactions) of the Brownian particles is denoted by $D$.
Each particle thus obeys the stochastic differential equation
\begin{equation}
\frac{d {\bf x}_i}{d t} = D\beta {\bf F}_i  + \sqrt{2D}\boldsymbol{\xi}_i\label{sde}
\end{equation}
where $\boldsymbol{\xi}_i$ is white noise in the Ito Stochastic Calculus with $\langle \xi_{i,\mu}(t) \xi_{j,\nu}(t')\rangle =\delta_{ij}\delta_{\mu\nu}\delta(t-t')$,  and ${\bf F}_i$ is the force acting on particle $i$, due to the potential $\Phi$. (Throughout, $i$ and $j$ label particles, while Greek indices label spacial components).
In other words, in the absence of $\Phi$, each particle performs isotropic Brownian motion.
\subsection{Obervables -- mean and fluctuating}\label{sec:Ob}
We summarize the arising observables in Table~\ref{table:1}. The  density operator, $\rho(\vct{x})=\sum_i\delta(\vct{x}-\vct{x}_i)$ \cite{HansenMcDonald} is the starting point for all considerations that follow. If averaged over the equilibrium distribution, one obtains the mean equilibrium density
\begin{align}
\left\langle \rho(\vct{x})\right\rangle= \left\langle\sum_i\delta(\vct{x}-\vct{x}_i)\right\rangle.\label{eq:barrho}
\end{align}
Here, we have introduced the {\it equilibrium} average $\left\langle \dots\right\rangle$, which, for the overdamped system is exactly given by (we introduce the phase space abbreviation $\Gamma\equiv\{{\vct{x}_i}\}$) 
\begin{align}\label{eq:av0}
\left\langle \dots\right\rangle=\frac{\int d\Gamma \dots   e^{- \beta\Phi(\Gamma)}}{\int d\Gamma   e^{- \beta\Phi(\Gamma)}}.
\end{align}
As noted above, for systems with infinite particle number, the grand canonical average  agrees with the canonical one given here.  
 We introduce density fluctuations,
\begin{align}
\phi(\vct{x})=\rho(\vct{x})-\langle\rho(\vct{x})\rangle.   \label{eq:phi}
\end{align}
This  quantity will be important for this manuscript. Such fluctuations are e.g.~characterized by their correlation function, which relates two points in space and in time $t$, (in the following we will sometimes suppress the arguments of $C$)
\begin{align}\label{eq:Cc}
C(\vct{x},\vct{x}',t-t')=\left\langle \phi(\vct{x},t)\phi(\vct{x}',t')\right\rangle.
\end{align} 
Due to the fact that we restrict to equilibrium fluctuations, $C$ is a function of $t-t'$ \cite{Risken}, but depends on both  $\vct{x}$ and $\vct{x}'$ in inhomogeneous systems. Computing $C$ is the main goal of the manuscript. If transformed to reciprocal Fourier (${\bf k}$-space), $\tilde C$ is the intermediate scattering function \cite{dhont}.  

For completeness, we also define the mean density in an out of equilibrium situation, i.e., $\langle\rho(\vct{x},t)\rangle^{\rm{neq}}$, and its average deviation from equilibrium 
\begin{align}\label{eq:noneq}
\delta\rho(\vct{x},t)=\langle\rho(\vct{x},t)\rangle^{\rm{neq}}-\left<\rho(\vct{x})\right>.
\end{align}
Note the difference to Eq.~\eqref{eq:phi}, which is for a stochastic fluctuation in an equilibrium system while Eq.~\eqref{eq:noneq}
 is an average deviation from the equilibrium average density for a perturbed system. 
\subsection{Effective Hamiltonian}\label{sec:eH}
Aiming at the correlation function $C$, we start by assuring that the equal time value of $C$ is found correctly. We thus introduce the following {\it effective} Hamiltonian, which is a functional of the fluctuating field $\phi$,
\begin{align}
\beta H= \frac{1}{2}\int d\vct{x} d\vct{y} \,\phi(\vct{x}) \frac{1}{\left\langle \phi(\vct{x}) \phi(\vct{y})\right\rangle}\phi(\vct{y}),\label{eq:H}
\end{align}  
where 
\begin{equation}
\frac{1}{\left\langle \phi(\vct{x}) \phi(\vct{y})\right\rangle} \equiv \left\langle \phi(\vct{x}) \phi(\vct{y})\right\rangle^{-1},
\end{equation}
is to be understood in the sense of inverse operators.
In the field theory description, the equilibrium average in Eq.~\eqref{eq:av0} is computed via the following functional  integral \cite{kardarbook, onukibook}  
\begin{align}
\left\langle \dots\right\rangle=\frac{\int {\cal D}\phi\dots  e^{- \beta H}}{\int {\cal D}\phi   e^{- \beta H}}.\label{eq:av}
\end{align} 
 $\cal{D}\phi$ denotes the measure of functional integration, which is most easily implemented  by  discretizing space
 or working with discrete Fourier transforms. As mentioned before, we point out that the Hamiltonian in Eq.~\eqref{eq:H} with Eq.~\eqref{eq:av} by construction correctly finds the static averages of $\phi$, up to quadratic order. Its average, $\left<\phi\right>=0$, as required from Eq.~\eqref{eq:phi}, and its variance is indeed, using Eq.~\eqref{eq:H},   
\begin{align}
\left\langle \phi(\vct{x}) \phi(\vct{y})\right\rangle=\frac{\int {\cal D}\phi  \phi(\vct{x})\phi(\vct{y})  e^{-\beta H}}{\int {\cal D}\phi   e^{-\beta H}}.\label{eq:var}
\end{align}
See e.g. Ref.~\cite{Altland} for useful identities regarding Gaussian functional integrals.
\subsection{Static correlations -- Theory of liquids}
The variance of $\phi$ in Eq.~\eqref{eq:var} is a well studied object in the theory of liquids, and can be expressed in terms of the so called direct pair correction function $c^{(2)}$ \cite{HansenMcDonald}, (this equation can be seen as one way of defining $c^{2}$)
\begin{align}
\frac{1}{\left\langle \phi(\vct{x}) \phi(\vct{y})\right\rangle}=\frac{1}{\left<\rho(\vct{x})\right>} \delta(\vct{x}-\vct{y})-c^{(2)}(\vct{x},\vct{y}).\label{eq:c2}
\end{align} 
Using this, we can make the Hamiltonian in Eq.~\eqref{eq:H} more explicit,
\begin{align}
\beta H=\frac{1}{2}\int  d\vct{x}  \frac{\phi(\vct{x})^2}{\left<\rho(\vct{x})\right>}- \frac{1}{2}\int d\vct{x} d\vct{y}  \phi(\vct{x}) c^{(2)}(\vct{x},\vct{y})\phi(\vct{y}).
\end{align} 
This shows the nature of the Hamiltonian: It has a local term, corresponding to the local compressibility of an ideal gas, and a nonlocal term, which is given by the direct correlation function. It is worth mentioning that the direct correlation function is a rather featureless function, and typically zero if $|\vct{x}-\vct{y}|$ is larger than the interaction range of the particles \cite{HansenMcDonald}. In contrast, the correlation function $\left\langle \phi(\vct{x}) \phi(\vct{y})\right\rangle$ can extend to larger distances, which, mathematically, is a consequence of taking the operator inverse. Physically, it is well known that correlations may reach further than interparticle interactions. 

We finish this subsection by introducing the short hand notation for the inverse of the static density correlation  
\begin{align}
\Delta(\vct{x},\vct{y})\equiv k_BT\left(\frac{1}{\left<\rho(\vct{x})\right>} \delta(\vct{x}-\vct{y})-c^{(2)}(\vct{x},\vct{y}).\label{eq:D}\right)
\end{align}
$\Delta(\vct{x},\vct{y})$ plays the role of an effective interaction potential between densities (we multiplied by $k_BT$ to obtain units of energy). In terms  of it, the Hamiltonian is finally,
\begin{align}
H=\frac{1}{2}\int  d\vct{x} d\vct{y}  \phi(\vct{x}) \Delta(\vct{x},\vct{y})\phi(\vct{y}).
\end{align}
\subsection{Equation of motion}\label{sec:eom}
While static equilibrium averages are determined via the Hamiltonian (with Eq.~\eqref{eq:av}), there is some freedom, or saying it differently, some lack of information, regarding the dynamics. The Hamiltonian in Eq.~\eqref{eq:H}  describes only a subset of degrees of freedom of the system. These degrees of freedom might however not capture all relevant features of the dynamics \cite{kardarbook, hohenberg}. The tools of classical mechanics are thus not applicable to deduce equations of motion from Eq.~\eqref{eq:H}. One possibility to overcome this problem is to resort to Langevin equations \cite{hohenberg,kardarbook}, which are based on deterministic (given by the explicit degrees of freedom) as well as stochastic forces (due to integrated degrees of freedom) \footnote{We note that in our setup, the starting point, Eq.~\eqref{sde} is already a Langevin equation. Introducing the effective Hamiltonian thus gives rise to an additional reduction of explicit degrees of freedom.}. The former may be written in terms of the driving force $\frac{\delta H}{\delta \phi}$, which gives the force due to deviations of $H$ from its minimum value. For our case, 
\begin{align}
\label{eq:d}
\beta\frac{\delta H}{\delta \phi}&=\frac{\phi(\vct{x})}{\langle\rho(\vct{x})\rangle} - \int d\vct{y} \,c^{(2)}(\vct{x},\vct{y})\phi(\vct{y})\\&\equiv\beta\Delta\phi.
\end{align}
In the second line, we used the short hand notation of Eq.~\eqref{eq:D}, and also a short hand notation for operator products, so that the second line contains an integration over the joint coordinate.  
Clearly,  $\left\langle\frac{\delta H}{\delta \phi}\right\rangle=0$, vanishing in  equilibrium. This force transforms into changes in $\phi$ with application of the operator $R=R(\vct{x},\vct{y})$, so that $R \frac{\delta H}{\delta \phi}$ involves an integral over the joint coordinate. $R$ involves among other things a mobility coefficient. $R$ having no time dependence, we have already restricted to a time local description for simplicity. Time non-local dynamics can be realized in this framework as well. Via the operator $R$, one can incorporate several types of dynamics, such as dynamics conserving the density, or not conserving it \cite{hohenberg,kardarbook,deangopinathan2010PRE}. We aim at dynamics of overdamped Brownian particles given by Eq.~\eqref{sde}, for  which -- justified a posteriori -- the proper choice for $R$ is,
\begin{align}
R=\frac{D}{k_BT}\nabla \cdot \langle\rho(\vct{x})\rangle\nabla\delta(\vct{x}-\vct{y}).\label{eq:R}
\end{align}
Note that, because $R$ is written as a divergence, local conservation of density is given. Indeed, the chosen $R$ in Eq.~\eqref{eq:R} is a version of the famous ``Model B'' \cite{hohenberg,kardarbook}. We thus write the following equation of motion, 
\begin{align}
\frac{\partial \phi}{\partial t}&= R \frac{\delta H}{\delta \phi}+ \nabla \cdot\sqrt{2D\langle\rho\rangle}\boldsymbol\eta(\vct{x},t)\label{eq:eomR},\\
&=\frac{D}{k_BT}\nabla \cdot \langle\rho(\vct{x})\rangle\nabla \frac{\delta H}{\delta \phi}+\nabla \cdot\sqrt{2D\langle\rho\rangle}\boldsymbol\eta(\vct{x},t)\label{eq:eom2}.
\end{align}
The included stochastic force is fixed through  the choice of the operator $R$. The field $\boldsymbol\eta(\vct{x},t)$ is  spatio-temporal vectorial white noise field with $\langle\boldsymbol\eta\rangle=0$, and whose components have the correlation function
\begin{align}
\left\langle\eta_\mu(\vct{x},t)\eta_\nu(\vct{y},t')\right\rangle= \delta_{\mu\nu}\delta(t-t')\delta(\vct{x}-\vct{y}).\label{eq:FDT}
\end{align}
The form of the last term in Eq.~\eqref{eq:eomR} and the variance in Eq.~\eqref{eq:FDT} ensures the validity of the fluctuation dissipation theorem \cite{deangopinathan2010PRE}, and makes sure that Eq.~\eqref{eq:eomR} finds the correct variance for $\phi$.   The explicit form of Eq.~\eqref{eq:eomR} reads 

\begin{align}
\frac{\partial \phi}{\partial t}=&D\nabla\cdot\left[\nabla \phi-\phi\nabla \log \langle\rho\rangle-\langle\rho\rangle\nabla\int d\vct{y} \,c^{(2)}(\vct{x},\vct{y})\phi(\vct{y})\right]\notag\\&+\nabla \cdot\sqrt{2D\langle\rho\rangle}{ \boldsymbol{\eta}}(\vct{x},t)
\end{align}
Examining this equation for the case of an ideal gas in the absence of an external potential, for which $c^{(2)}=0$ and $\langle\rho\rangle$ spatially constant, we obtain, as expected, a diffusion equation with a conservative noise term,
\begin{align}
\frac{\partial \phi}{\partial t} = D\nabla^2 \phi+\nabla \cdot\sqrt{2D\langle\rho\rangle}\boldsymbol\eta.
\label{eq:9}
\end{align}
While the above equation contains the correct diffusion term for the ideal gas, it yields by construction Gaussian density fluctuations (because it is implied that the noise correlation in Eq.~\eqref{eq:FDT} is Gaussian, so that higher order correlations of ${\bf\eta}$ can be factorized). As a sidenote, we remark that, interestingly, even for an ideal gas, the distribution of $\phi$ is nontrivial and is in fact Poissonian \cite{velenich08}. This difference in underlying statistics only shows up for higher point correlation functions, so that the present theory is exact for the two point correlations in the case of ideal gas. This will be demonstrated  at the end of subsection \ref{sec:genres}.

Together with Eq.~\eqref{eq:d}, Eq.~\eqref{eq:eom2} gives a closed form for the dynamics of the system, which is our first main result. This dynamics is chosen to yield exact equilibrium correlation functions but also gives the exact time-correlation function for non-interacting Brownian particles. In the following sections, we will investigate the properties of the dynamics proposed here in more detail.

\section{Time dependent correlation function}\label{sec:tdc}
In this section, we  finally  compute and analyze the resulting approximative form of the time dependent equilibrium correlation function, as defined in Eq.~\eqref{eq:Cc}, as following from Eq.~\eqref{eq:eom2}.  
\subsection{General result}\label{sec:genres}
We start by writing the Langevin equation, Eq.~\eqref{eq:eom2} in a shorter form, using Eq.~\eqref{eq:D}, 
\begin{align}
\notag\frac{\partial \phi}{\partial t}&=\frac{D}{k_BT}\nabla \cdot \langle\rho(\vct{x})\rangle\nabla \frac{\delta H}{\delta \phi} +\nabla \cdot\sqrt{2D\langle\rho\rangle}\boldsymbol\eta,\\
&= R\Delta\phi+\nabla \cdot\sqrt{2D\langle\rho\rangle}\boldsymbol\eta,\label{eq:eom3}
\end{align} 
Eq.~\eqref{eq:eom3} can then be easily solved for the correlation function, from its general solution, (see, e.g., Ref.~\cite{deangopinathan2010PRE}),
\begin{align} 
\phi(t)=\phi(t_0)+\int_{t_0}^t ds \,e^{(t-s)R\Delta} \nabla \cdot\sqrt{2D\langle\rho\rangle}\boldsymbol\eta(s). 
\end{align}
The average of  $\phi(t)\phi(t')$ over the noise contains then several terms, including terms depending on the initial value at $t_0$. Aiming at the equilibrium correlation function, we let $t$ and $t'$ formally go to infinity, and obtain the steady equilibrium part, which depends only on $t-t'$ (recall that, as before, $\langle\dots \rangle$ denotes an average in equilibrium),
\begin{align} 
C=\left\langle \phi(\vct{x},t)\phi(\vct{x}',t')\right\rangle=\frac{k_BT}{\Delta}e^{|t-t'|\Delta R}. \label{eq:C}
\end{align}  
The correlation function is generally not an exponential in time, because $R$ and $\Delta$ are operators. Eq.~\eqref{eq:C} is our second main result.

We can now show that Eq.~\eqref{eq:C} is exact for non-interaction particles. To see this it is best to work in Fourier space where the density operator $\rho$
takes the form, for $N$ particles,
\begin{align}
\tilde \rho({\bf k}) = \sum_{i=1}^N \exp(-i{\bf k}\cdot{\bf x}_i)
\end{align}
where  ${\bf x}_i$ obeys Eq.~\eqref{sde} with $\bf F=0$.  
The ensemble average in this free gas is over the trajectories of the Brownian motions  ${\bf \xi}_i$. 
The average of $\tilde \rho({\bf k}) $ is given by
\begin{align}
\langle \tilde \rho({\bf k})\rangle = (2\pi)^d\delta({\bf k})\left\langle \rho\right\rangle,  
\end{align}
with here $N/V = \left\langle \rho\right\rangle$ the uniform bulk density. A simple computation shows that the two point correlation function of the fluctuations $\phi$ at different times is given for large $N$ by
\begin{equation}\label{eq:ig}
\langle \tilde \phi({\bf k},t)\tilde \phi({\bf k}',0)\rangle = (2\pi)^d\delta({\bf k}+{\bf k}')\left\langle \rho\right\rangle\exp(-D k^2t).
\end{equation}
Transforming Eq.~\eqref{eq:C} (derived from Eq.~\eqref{eq:9}) to Fourier space, the agreement to the independently obtained Eq.~\eqref{eq:ig} can easily be verified.  

\subsection{Comparing to exact solution for short times}  \label{sec:ste}
For small values of time $t-t'$, we expand Eq.~\eqref{eq:C},
\begin{align} 
\left\langle \phi(\vct{x},t)\phi(\vct{x}',t')\right\rangle=\frac{k_BT}{\Delta}\left (1+|t-t'|\Delta R+\dots\right). \label{eq:st1}
\end{align}
The dots represent higher order terms in $t-t'$.
We shall now compare this result to the exact one for Eq.~\eqref{sde}. (Recall that we assume that canonical and grand canonical systems are equivalent). For this, we use the Smoluchowski equation corresponding to the set of stochastic equations, Eq.~\eqref{sde}. The Smoluchowski equation is a partial differential equation for the distribution $\Psi(\Gamma,t)$, which is a function of phase space $\Gamma$ \cite{Risken},
\begin{equation}\label{eq:Smo}
\frac{\partial}{\partial t}\Psi= \Omega\Psi.
\end{equation} 
$\Omega=D\sum_{i}\boldsymbol{\partial}_i\cdot[ \boldsymbol{\partial}_i-\beta{\bf F}_i]$ is the Smoluchowski operator. ${\bf F}_i$ is, as in Eq.~\eqref{sde}, the force acting on particle $i$. The equilibrium time correlation function for density is then written \cite{fuchs}, 
\begin{align}\label{eq:SM}
\left\langle \phi(\vct{x},t)\phi(\vct{x}',t')\right\rangle=\int d\Gamma  \phi(\vct{x}) e^{ |t-t'|\Omega}  \phi(\vct{x}') \Psi_e(\Gamma).
\end{align}
Here, $\Psi_e$ is the equilibrium distribution. For short times, Eq.~\eqref{eq:SM} is expanded,
\begin{align}
\left\langle \phi(\vct{x},t)\phi(\vct{x}',t')\right\rangle=
\left\langle \phi(\vct{x})\phi(\vct{x}')\right\rangle+\notag\\ |t-t'|\int d\Gamma  \phi(\vct{x} )\Omega  \phi(\vct{x}') \Psi_e(\Gamma)+\dots
\label{eq:st}
\end{align}
$\Psi_e$ being  the  Boltzmann distribution, one has  ${\boldsymbol \partial_i}\Psi_e =\beta\vct{F}_i\Psi_e$ \cite{Risken,fuchs}. 
Using this we can rewrite the second term in Eq.~\eqref{eq:st} by use of partial integrations (Einstein summation convention is implied),   
\begin{align}
\int d\Gamma  \phi(\vct{x} )\Omega  \phi(\vct{x}') \Psi_e(\Gamma)= -D\int d\Gamma (\partial_i \phi(\vct{x}))(\partial_i   \phi(\vct{x'}))\Psi_e.\label{eq:st4}
\end{align}
We  now employ the definition of  $\phi(\vct{x})=\sum_i \delta(\vct{x}-\vct{x}_i) - \langle\rho(\vct{x})\rangle$, noticing that the mean density vanishes when plugged into Eq.~\eqref{eq:st4}: It does not depend on phase space and $\partial_i$ yields zero. With  $\partial_i \delta(\vct{x}-\vct{x}_i)=-\partial_x \delta(\vct{x}-\vct{x}_i)$, we get
\begin{align}
-D\langle  (\partial_i \phi(\vct{x}))(\partial_i   \phi(\vct{x}'))\rangle \notag\\=\notag-D\sum_i\nabla \langle \delta(\vct{x}-\vct{x}_i) \delta(\vct{x}'-\vct{x}_i) \rangle \overleftarrow{\nabla'}\\=-D\sum_i\nabla \delta(\vct{x}-\vct{x'}) \langle\rho\rangle (\vct{x}') \overleftarrow\nabla'=k_BT R(\vct{x},\vct{x}')\label{eq:st2}.
\end{align}
Where we have identified the operator $R$ from Eq.~\eqref{eq:R}. We thus have the exact solution for short times,
\begin{align} 
\left\langle \phi(t)\phi(t')\right\rangle=\frac{k_BT}{\Delta}\left (1+|t-t'|\Delta R+\dots\right). \label{eq:st3}
\end{align}
Comparison with Eq.~\eqref{eq:st1} reveals that Eq.~\eqref{eq:C} agrees with the exact solution of the Smoluchowski equation for short times, i.e., including the term linear in time. This linear term has been discussed in terms of the ''initial decay rate'' \cite{Fuchs09}, or in terms of a wavevector dependend diffusivity \cite{Pusey,Cichocki91,Cichocki93}. The current formulation agrees with these. 

\section{Relaxation to equilibrium -- Agreement with DDFT}\label{sec:DDFT}
While in the previous section, we compared the stochastic equation proposed here (Eq.~\eqref{eq:eom2}) to the exact Smoluchowski equation, in this section, we aim to demonstrate another equivalence: Near equilibrium, the relaxation dynamics of Eq.~\eqref{eq:eom2}  agrees exactly with the corresponding result of DDFT. This will be seen by studying the relaxation of a system which is initially out of equilibrium.
\subsection{Mean relaxation to equilibrium from Eq.~\eqref{eq:eom2}}     
Let us assume, the system is in an initial situation {\it out} of equilibrium, so that the mean density deviates from the equilibrium one, and we define as in Tab.~\ref{table:1},
\begin{align}
\langle\rho(\vct{x})\rangle^{\it neq}=\langle\rho(\vct{x})\rangle+\delta\rho(\vct{x}).
\end{align}  
If $\delta\rho(\vct{x})$ is small, we can use Eq.~\eqref{eq:eom2} to compute the relaxation of $\delta\rho$ to zero ($\delta\rho(\vct{x})$ must be small because Eq.~\eqref{eq:eom2} is linear). Therefore, we replace $\phi$ in Eq.~\eqref{eq:eom2} by $\delta\rho$, and remove the noise term, as it vanishes when taking the mean of the equation.  We obtain the following equation which is linear in $\delta\rho$,
\begin{align}
\frac{\partial \delta\rho}{\partial t} = &D\nabla\cdot\biggl[\nabla \delta\rho-\delta\rho\nabla \log \langle\rho\rangle\notag\\&\left.-\langle\rho\rangle\nabla\int d\vct{y} \,c^{(2)}(\vct{x},\vct{y})\delta \rho(\vct{y})\right] .\label{eq:8}
\end{align}  
In the next subsection, we will compute the analogous equation from DDFT, and demonstrate the agreement.   
\subsection{DDFT expanded near equilibrium}
Quoting the equation of motion of dynamical density functional theory for Brownian particles in an external potential $U$ \cite{archer}, one has
\begin{align}
\frac{\partial \delta\rho}{\partial t}=&D\nabla\cdot\biggl[\nabla(\langle\rho\rangle+\delta\rho)+\beta(\langle\rho\rangle+\delta\rho)\nabla U\notag\\&\left.+\beta(\langle\rho\rangle+\delta\rho)\nabla \frac{\delta {\cal F}^{ex}}{\delta \rho}\right].\label{eq:DDFT}
\end{align}
${\cal F}^{ex}$ is the so called excess free energy functional. 
This is a well known and well studied equation, which is an approximative solution of the Smoluchowski equation, Eq.~\eqref{eq:Smo}. It has been successfully used in many scenarios to describe the dynamics of interacting Brownian particles \cite{archer}. We now expand this equation for small values of $\delta\rho$, as in Eq.~\eqref{eq:8}. We first note that several terms cancel, as the time derivative must vanish in equilibrium. Specifically (note that even the term in the square brackets vanishes), 
\begin{align}
0=\nabla\cdot\left[\nabla\langle\rho\rangle+\beta\langle\rho\rangle\nabla U+\beta\langle\rho\rangle\nabla \left.\frac{\delta {\cal F}^{ex}}{\delta \rho}\right|_{\rho=\langle\rho\rangle}\right].\label{eq:z}
\end{align}
Furthermore, for small $\delta\rho$, we expand the last term in Eq.~\eqref{eq:DDFT} in a functional Taylor series,
\begin{align}
\frac{\delta {\cal F}^{ex}}{\delta \rho(\vct{x})}= \left.\frac{\delta {\cal F}^{ex}}{\delta \rho(\vct{x})}\right|_{\rho=\langle\rho\rangle}+ \int d\vct{y} \,\left.\frac{\delta {\cal F}^{ex}}{\delta \rho(\vct{x})\delta \rho(\vct{y})}\right|_{\rho=\langle\rho\rangle} \delta\rho(\vct{y})\notag\\+\mathcal{O}(\delta\rho^2).\label{eq:Taylor}
\end{align}
It is now important to note that the involved Taylor coefficient equals, by definition, the direct correlation function $c^{(2)}$ \cite{HansenMcDonald},
\begin{align}
-\beta\left.\frac{\delta{\cal F}^{ex}}{\delta\rho(\vct{x})\delta\rho(\vct{y})}\right|_{\rho=\langle\rho\rangle}= c^{(2)}(\vct{x},\vct{y})\label{eq:Taylorc}.
\end{align}
Another useful relation is the formal exact result for the equilibrium mean density, which is given by \cite{HansenMcDonald},
\begin{align}
\langle\rho\rangle=z \exp\left[-\beta U-\left.\frac{\beta\delta {\cal F}^{ex}}{\delta \rho(\vct{x})}\right|_{\rho=\langle\rho\rangle}\right],
\end{align}
with the (in the following irrelevant) normalization $z$. With this equation, one can write
\begin{align}
k_BT\nabla\log \langle\rho\rangle=-\nabla U-\nabla \left.\frac{\delta {\cal F}^{ex}}{\delta \rho(\vct{x})}\right|_{\rho=\langle\rho\rangle}.\label{eq:logr}
\end{align}
We finally obtain for the expansion of Eq.~\eqref{eq:DDFT} linear in $\delta \rho$,
\begin{align}
\frac{\partial \delta\rho}{\partial t} = &D\nabla\cdot\biggl[\nabla \delta\rho-\delta\rho\nabla \log \langle\rho\rangle\notag\\&\left.-\langle\rho\rangle\nabla\int d\vct{y} \,c^{(2)}(\vct{x},\vct{y})\delta \rho(\vct{y})\right]
\label{eq:8_1}
\end{align}
which is identical to Eq.~\eqref{eq:8}. We have thus shown that the new equation, Eq.~\eqref{eq:eom2} is in agreement with DDFT for small deviations from equilibrium. This demonstrates a connection to the framework of Ref.~\cite{Archer04b}, without, however, an obvious direct equivalence.
\section{Comparison to the exact stochastic equation}\label{sec:SDE}
Starting from the set of stochastic equations in Eq.~\eqref{sde}, the following exact stochastic equation of motion is found for the  density operator $\rho$ (see Table~\ref{table:1}) \cite{Dean96},     
\begin{equation}
\frac{\partial}{\partial t} \rho({\bf x}) =D \nabla \cdot  \rho\nabla \frac{\delta \beta {\cal E}}{\delta \rho({\bf x})} +\nabla\cdot \sqrt{2D\rho}{\boldsymbol \eta}({\bf x},t).\label{dk1}
\end{equation}
Here, the noise $\eta$ is distributed as in Eq.~\eqref{eq:FDT}, and {$\cal E$} is the energy functional
\begin{align}\label{eq:pot}
{\cal E}=&k_BT\int d{\bf x} \rho({\bf x}) \ln(\rho({\bf x})) \notag\\&+ \frac{1}{2}\int d{\bf x}d{\bf y} \rho({\bf x}) V({\bf x}-{\bf y}) \rho({\bf y})\notag\\&+\int d{\bf x} \rho({\bf x})U(\vct{x}).
\end{align}
We have as before split the potential into an interaction part $V$, and an external part $U$. Note the difference of Eq.~\eqref{eq:pot} compared to the free energy functional of DFT \cite{HansenMcDonald, Dean96,Archer04b}.
Attempting to linearize Eq.~\eqref{dk1} in the fluctuations $\phi$, the first natural choice is to replace the density operator appearing in the noise term by its equilibrium average, i.e., $\sqrt{2D\rho}{\boldsymbol \eta}({\bf x},t)\approx\sqrt{2D\langle\rho\rangle}{\boldsymbol\eta}({\bf x},t)$. Interestingly, in order to keep detailed balance, this choice implies that also the density operator on the right hand side of Eq.~\eqref{dk1} must be replaced by its mean, i.e., the resulting approximate, consistent linear equation reads
\begin{equation}
\frac{\partial}{\partial t} \phi({\bf x}) =D \nabla \cdot  \langle\rho\rangle\nabla \frac{\delta \beta {\cal E}'}{\delta \phi({\bf x})} +\nabla\cdot \sqrt{2D\langle\rho\rangle}{\bf \boldsymbol\eta}({\bf x},t).\label{dk2}
\end{equation}  
Furthermore, we note Eq.~\eqref{dk2} yields the correct result for the variance of $\phi$ in equilibrium, $\langle \phi(\vct{x})\phi(\vct{y})\rangle$, if the functional ${\cal E}'$ coincides with $H$ of Eq.~\eqref{eq:H}, ${\cal E}'=H$.

It is thus interesting to note that after pre-averaging the noise, Eq.~\eqref{eq:eom2} appears to be the only consistent, linear equation for $\phi$. It has been recently shown that linearizing the interaction term in Eq.~\eqref{dk1} about the mean bulk density, while using the mean bulk density in the noise term, leads to an analytically soluble theory in 
the bulk which recovers the random phase approximation for the equal time correlation functions \cite{dean14, demery14,lu15,dean16,demery16}, notably this means that Debye-H\"uckel theory is obtained for Brownian electrolytes. The approach has been applied to a variety of driven and out of equilibrium systems. In particular it is  capable of reproducing the full Onsager theory of electrolyte  conductivity, both the Ohmic linear response regime and the 
first Wien effect regime where the conductivity is enhanced by the electric field \cite{demery16}. While the random phase approximation is valid only for weak interactions or high temperatures, the approach here should allow the study of systems with  form instance hard core interactions, relevant for ionic liquids, both in the bulk and under confinement.

\section{Summary}\label{sec:Su}
We have derived an effective field theory for simple liquids, which allows computation of dynamical correlation functions of the density. The result for the dynamical correlation function is approximate, but exact for small times. The described dynamics also agrees exactly at all times  with dynamical density functional theory. 
Future work will apply this theory to study the intermediate scattering function in confinement (where recent experimental findings exist \cite{Nygard16}). It will also be used to find the local viscosity near surfaces and compare to previous theoretical approaches for bulk \cite{Johannes_Joe_2013} and confinement \cite{,AerovKrugerJCP2014,aerov2015}. For this, an expression for the stress tensor in this theory must be derived. Then this theory may also be used to investigate out of equilibrium Casimir forces in model B as in Ref.~\cite{Rohwer17}, however including effects of finite particle size.

\begin{acknowledgments}
We thank C. Rohwer for helpful discussions. This research was supported by the DFG grant No. KR 3844/2-1 and the ANR project FISICS.
\end{acknowledgments}


\begin{thebibliography}{71}%
\makeatletter
\providecommand \@ifxundefined [1]{%
 \@ifx{#1\undefined}
}%
\providecommand \@ifnum [1]{%
 \ifnum #1\expandafter \@firstoftwo
 \else \expandafter \@secondoftwo
 \fi
}%
\providecommand \@ifx [1]{%
 \ifx #1\expandafter \@firstoftwo
 \else \expandafter \@secondoftwo
 \fi
}%
\providecommand \natexlab [1]{#1}%
\providecommand \enquote  [1]{``#1''}%
\providecommand \bibnamefont  [1]{#1}%
\providecommand \bibfnamefont [1]{#1}%
\providecommand \citenamefont [1]{#1}%
\providecommand \href@noop [0]{\@secondoftwo}%
\providecommand \href [0]{\begingroup \@sanitize@url \@href}%
\providecommand \@href[1]{\@@startlink{#1}\@@href}%
\providecommand \@@href[1]{\endgroup#1\@@endlink}%
\providecommand \@sanitize@url [0]{\catcode `\\12\catcode `\$12\catcode
  `\&12\catcode `\#12\catcode `\^12\catcode `\_12\catcode `\%12\relax}%
\providecommand \@@startlink[1]{}%
\providecommand \@@endlink[0]{}%
\providecommand \url  [0]{\begingroup\@sanitize@url \@url }%
\providecommand \@url [1]{\endgroup\@href {#1}{\urlprefix }}%
\providecommand \urlprefix  [0]{URL }%
\providecommand \Eprint [0]{\href }%
\providecommand \doibase [0]{http://dx.doi.org/}%
\providecommand \selectlanguage [0]{\@gobble}%
\providecommand \bibinfo  [0]{\@secondoftwo}%
\providecommand \bibfield  [0]{\@secondoftwo}%
\providecommand \translation [1]{[#1]}%
\providecommand \BibitemOpen [0]{}%
\providecommand \bibitemStop [0]{}%
\providecommand \bibitemNoStop [0]{.\EOS\space}%
\providecommand \EOS [0]{\spacefactor3000\relax}%
\providecommand \BibitemShut  [1]{\csname bibitem#1\endcsname}%
\let\auto@bib@innerbib\@empty
\bibitem [{\citenamefont {Hansen}\ and\ \citenamefont
  {McDonald}(2009)}]{HansenMcDonald}%
  \BibitemOpen
  \bibfield  {author} {\bibinfo {author} {\bibfnamefont {J.-P.}\ \bibnamefont
  {Hansen}}\ and\ \bibinfo {author} {\bibfnamefont {I.}~\bibnamefont
  {McDonald}},\ }\href@noop {} {\emph {\bibinfo {title} {Theory of simple
  liquids}}}\ (\bibinfo  {publisher} {Academic Press},\ \bibinfo {year}
  {2009})\BibitemShut {NoStop}%
\bibitem [{\citenamefont {Evans}(1992)}]{bob_review}%
  \BibitemOpen
  \bibfield  {author} {\bibinfo {author} {\bibfnamefont {R.}~\bibnamefont
  {Evans}},\ }\href@noop {} {\emph {\bibinfo {title} {Fundamentals of
  inhomogeneous fluids}}}\ (\bibinfo  {publisher} {Dekker},\ \bibinfo {address}
  {New York},\ \bibinfo {year} {1992})\BibitemShut {NoStop}%
\bibitem [{\citenamefont {G\"otze}(2008)}]{Goetze}%
  \BibitemOpen
  \bibfield  {author} {\bibinfo {author} {\bibfnamefont {W.}~\bibnamefont
  {G\"otze}},\ }\href@noop {} {\emph {\bibinfo {title} {Complex Dynamics of
  Glass-Forming Liquids: A Mode-Coupling Theory}}}\ (\bibinfo  {publisher}
  {Oxford University Press},\ \bibinfo {address} {Oxford},\ \bibinfo {year}
  {2008})\BibitemShut {NoStop}%
\bibitem [{\citenamefont {Roth}(2010)}]{roth_review}%
  \BibitemOpen
  \bibfield  {author} {\bibinfo {author} {\bibfnamefont {R.}~\bibnamefont
  {Roth}},\ }\href@noop {} {\bibfield  {journal} {\bibinfo  {journal}
  {J.Phys.:Condens.Matter}\ }\textbf {\bibinfo {volume} {22}},\ \bibinfo
  {pages} {063102} (\bibinfo {year} {2010})}\BibitemShut {NoStop}%
\bibitem [{\citenamefont {Maxwell}(1866)}]{Maxwell_viscosity}%
  \BibitemOpen
  \bibfield  {author} {\bibinfo {author} {\bibfnamefont {J.}~\bibnamefont
  {Maxwell}},\ }\href@noop {} {\bibfield  {journal} {\bibinfo  {journal}
  {Philosophical Transactions of the Royal Society of London}\ }\textbf
  {\bibinfo {volume} {156}},\ \bibinfo {pages} {249} (\bibinfo {year}
  {1866})}\BibitemShut {NoStop}%
\bibitem [{\citenamefont {Viswanath}\ \emph {et~al.}(2007)\citenamefont
  {Viswanath} \emph {et~al.}}]{Viscosity_of_Liquids}%
  \BibitemOpen
  \bibfield  {author} {\bibinfo {author} {\bibfnamefont {D.~S.}\ \bibnamefont
  {Viswanath}} \emph {et~al.},\ }\href@noop {} {\emph {\bibinfo {title}
  {Viscosity of Liquids}}}\ (\bibinfo  {publisher} {Springer},\ \bibinfo
  {address} {The Netherlands},\ \bibinfo {year} {2007})\BibitemShut {NoStop}%
\bibitem [{\citenamefont {Elliott}\ \emph {et~al.}(2014)\citenamefont {Elliott}
  \emph {et~al.}}]{ElliottPRL2014ViscFermGasMeasurement}%
  \BibitemOpen
  \bibfield  {author} {\bibinfo {author} {\bibfnamefont {E.}~\bibnamefont
  {Elliott}} \emph {et~al.},\ }\href@noop {} {\bibfield  {journal} {\bibinfo
  {journal} {Phys. Rev. Lett.}\ }\textbf {\bibinfo {volume} {113}},\ \bibinfo
  {pages} {020406} (\bibinfo {year} {2014})}\BibitemShut {NoStop}%
\bibitem [{\citenamefont {Green}(1952)}]{Green}%
  \BibitemOpen
  \bibfield  {author} {\bibinfo {author} {\bibfnamefont {M.}~\bibnamefont
  {Green}},\ }\href@noop {} {\bibfield  {journal} {\bibinfo  {journal} {J.
  Chem. Phys.}\ }\textbf {\bibinfo {volume} {20}},\ \bibinfo {pages} {1281}
  (\bibinfo {year} {1952})}\BibitemShut {NoStop}%
\bibitem [{\citenamefont {Kubo}(1957)}]{Kubo}%
  \BibitemOpen
  \bibfield  {author} {\bibinfo {author} {\bibfnamefont {R.}~\bibnamefont
  {Kubo}},\ }\href@noop {} {\bibfield  {journal} {\bibinfo  {journal} {J. Phys.
  Soc. Jpn.}\ }\textbf {\bibinfo {volume} {12}},\ \bibinfo {pages} {570}
  (\bibinfo {year} {1957})}\BibitemShut {NoStop}%
\bibitem [{\citenamefont {Dhont}(1996)}]{dhont}%
  \BibitemOpen
  \bibfield  {author} {\bibinfo {author} {\bibfnamefont {J.~K.~G.}\
  \bibnamefont {Dhont}},\ }\href@noop {} {\emph {\bibinfo {title} {An
  Introduction to Dynamics of Colloids}}}\ (\bibinfo  {publisher} {Elsevier
  science},\ \bibinfo {address} {Amsterdam},\ \bibinfo {year}
  {1996})\BibitemShut {NoStop}%
\bibitem [{\citenamefont {Risken}(1984)}]{Risken}%
  \BibitemOpen
  \bibfield  {author} {\bibinfo {author} {\bibfnamefont {H.}~\bibnamefont
  {Risken}},\ }\href@noop {} {\emph {\bibinfo {title} {The Fokker-Planck
  Equation}}}\ (\bibinfo  {publisher} {Springer},\ \bibinfo {address}
  {Berlin},\ \bibinfo {year} {1984})\BibitemShut {NoStop}%
\bibitem [{\citenamefont {Kreuzer}(1981)}]{Kreuzer}%
  \BibitemOpen
  \bibfield  {author} {\bibinfo {author} {\bibfnamefont {H.~J.}\ \bibnamefont
  {Kreuzer}},\ }\href@noop {} {\emph {\bibinfo {title} {Nonequilibrium
  thermodynamics and its statistical foundations}}}\ (\bibinfo  {publisher}
  {Clarendon press},\ \bibinfo {address} {Oxford},\ \bibinfo {year}
  {1981})\BibitemShut {NoStop}%
\bibitem [{\citenamefont {da~C.~Andrade}(1934)}]{Andrade}%
  \BibitemOpen
  \bibfield  {author} {\bibinfo {author} {\bibfnamefont {E.}~\bibnamefont
  {da~C.~Andrade}},\ }\href@noop {} {\bibfield  {journal} {\bibinfo  {journal}
  {London Edinb. Dub. Philos. Mag. J. Sci.}\ }\textbf {\bibinfo {volume}
  {17(112)}},\ \bibinfo {pages} {497} (\bibinfo {year} {1934})}\BibitemShut
  {NoStop}%
\bibitem [{\citenamefont {Pusey}\ and\ \citenamefont {Tough}(1986)}]{Pusey}%
  \BibitemOpen
  \bibfield  {author} {\bibinfo {author} {\bibfnamefont {P.}~\bibnamefont
  {Pusey}}\ and\ \bibinfo {author} {\bibfnamefont {R.}~\bibnamefont {Tough}},\
  }\href@noop {} {\emph {\bibinfo {title} {In: Dynamic Light Scattering, Edited
  by R. Pecora}}}\ (\bibinfo  {publisher} {Plenum Press},\ \bibinfo {address}
  {New York and London},\ \bibinfo {year} {1986})\ p.~\bibinfo {pages}
  {85}\BibitemShut {NoStop}%
\bibitem [{\citenamefont {Lang}\ \emph {et~al.}(2010)\citenamefont {Lang},
  \citenamefont {Bo\ifmmode~\mbox{\c{t}}\else \c{t}\fi{}an}, \citenamefont
  {Oettel}, \citenamefont {Hajnal}, \citenamefont {Franosch},\ and\
  \citenamefont {Schilling}}]{Lang10}%
  \BibitemOpen
  \bibfield  {author} {\bibinfo {author} {\bibfnamefont {S.}~\bibnamefont
  {Lang}}, \bibinfo {author} {\bibfnamefont {V.}~\bibnamefont
  {Bo\ifmmode~\mbox{\c{t}}\else \c{t}\fi{}an}}, \bibinfo {author}
  {\bibfnamefont {M.}~\bibnamefont {Oettel}}, \bibinfo {author} {\bibfnamefont
  {D.}~\bibnamefont {Hajnal}}, \bibinfo {author} {\bibfnamefont
  {T.}~\bibnamefont {Franosch}}, \ and\ \bibinfo {author} {\bibfnamefont
  {R.}~\bibnamefont {Schilling}},\ }\href@noop {} {\bibfield  {journal}
  {\bibinfo  {journal} {Phys. Rev. Lett.}\ }\textbf {\bibinfo {volume} {105}},\
  \bibinfo {pages} {125701} (\bibinfo {year} {2010})}\BibitemShut {NoStop}%
\bibitem [{\citenamefont {Lang}\ \emph {et~al.}(2012)\citenamefont {Lang},
  \citenamefont {Schilling}, \citenamefont {Krakoviack},\ and\ \citenamefont
  {Franosch}}]{Lang12}%
  \BibitemOpen
  \bibfield  {author} {\bibinfo {author} {\bibfnamefont {S.}~\bibnamefont
  {Lang}}, \bibinfo {author} {\bibfnamefont {R.}~\bibnamefont {Schilling}},
  \bibinfo {author} {\bibfnamefont {V.}~\bibnamefont {Krakoviack}}, \ and\
  \bibinfo {author} {\bibfnamefont {T.}~\bibnamefont {Franosch}},\ }\href@noop
  {} {\bibfield  {journal} {\bibinfo  {journal} {Phys. Rev. E}\ }\textbf
  {\bibinfo {volume} {86}},\ \bibinfo {pages} {021502} (\bibinfo {year}
  {2012})}\BibitemShut {NoStop}%
\bibitem [{\citenamefont {Lang}\ \emph {et~al.}(2014)\citenamefont {Lang},
  \citenamefont {Schilling},\ and\ \citenamefont {Franosch}}]{Lang14}%
  \BibitemOpen
  \bibfield  {author} {\bibinfo {author} {\bibfnamefont {S.}~\bibnamefont
  {Lang}}, \bibinfo {author} {\bibfnamefont {R.}~\bibnamefont {Schilling}}, \
  and\ \bibinfo {author} {\bibfnamefont {T.}~\bibnamefont {Franosch}},\
  }\href@noop {} {\bibfield  {journal} {\bibinfo  {journal} {Phys. Rev. E}\
  }\textbf {\bibinfo {volume} {90}},\ \bibinfo {pages} {062126} (\bibinfo
  {year} {2014})}\BibitemShut {NoStop}%
\bibitem [{\citenamefont {Lang}\ and\ \citenamefont
  {Franosch}(2014)}]{Lang14_2}%
  \BibitemOpen
  \bibfield  {author} {\bibinfo {author} {\bibfnamefont {S.}~\bibnamefont
  {Lang}}\ and\ \bibinfo {author} {\bibfnamefont {T.}~\bibnamefont
  {Franosch}},\ }\href@noop {} {\bibfield  {journal} {\bibinfo  {journal}
  {Phys. Rev. E}\ }\textbf {\bibinfo {volume} {89}},\ \bibinfo {pages} {062122}
  (\bibinfo {year} {2014})}\BibitemShut {NoStop}%
\bibitem [{\citenamefont {Marconi}\ and\ \citenamefont
  {Tarazona}(1999)}]{Marconi99}%
  \BibitemOpen
  \bibfield  {author} {\bibinfo {author} {\bibfnamefont {U.~M.~B.}\
  \bibnamefont {Marconi}}\ and\ \bibinfo {author} {\bibfnamefont
  {P.}~\bibnamefont {Tarazona}},\ }\href@noop {} {\bibfield  {journal}
  {\bibinfo  {journal} {J. Chem. Phys.}\ }\textbf {\bibinfo {volume} {110}},\
  \bibinfo {pages} {8032} (\bibinfo {year} {1999})}\BibitemShut {NoStop}%
\bibitem [{\citenamefont {Archer}\ and\ \citenamefont {Evans}(2004)}]{archer}%
  \BibitemOpen
  \bibfield  {author} {\bibinfo {author} {\bibfnamefont {A.~J.}\ \bibnamefont
  {Archer}}\ and\ \bibinfo {author} {\bibfnamefont {R.}~\bibnamefont {Evans}},\
  }\href@noop {} {\bibfield  {journal} {\bibinfo  {journal} {J. Chem. Phys.}\
  }\textbf {\bibinfo {volume} {121}},\ \bibinfo {pages} {4246} (\bibinfo {year}
  {2004})}\BibitemShut {NoStop}%
\bibitem [{\citenamefont {Rex}\ and\ \citenamefont
  {L\"owen}(2008)}]{RexLoewenPRL2008}%
  \BibitemOpen
  \bibfield  {author} {\bibinfo {author} {\bibfnamefont {M.}~\bibnamefont
  {Rex}}\ and\ \bibinfo {author} {\bibfnamefont {H.}~\bibnamefont {L\"owen}},\
  }\href@noop {} {\bibfield  {journal} {\bibinfo  {journal} {Phys. Rev. Lett.}\
  }\textbf {\bibinfo {volume} {101}},\ \bibinfo {pages} {148302} (\bibinfo
  {year} {2008})}\BibitemShut {NoStop}%
\bibitem [{\citenamefont {Penna}\ \emph {et~al.}(2003)\citenamefont {Penna},
  \citenamefont {Dzubiella},\ and\ \citenamefont {Tarazona}}]{penna}%
  \BibitemOpen
  \bibfield  {author} {\bibinfo {author} {\bibfnamefont {F.}~\bibnamefont
  {Penna}}, \bibinfo {author} {\bibfnamefont {J.}~\bibnamefont {Dzubiella}}, \
  and\ \bibinfo {author} {\bibfnamefont {P.}~\bibnamefont {Tarazona}},\
  }\href@noop {} {\bibfield  {journal} {\bibinfo  {journal} {Physical Review
  E}\ } (\bibinfo {year} {2003})}\BibitemShut {NoStop}%
\bibitem [{\citenamefont {Rauscher}\ \emph {et~al.}(2007)\citenamefont
  {Rauscher}, \citenamefont {Dominguez}, \citenamefont {Kr\"uger},\ and\
  \citenamefont {Penna}}]{rauscher2}%
  \BibitemOpen
  \bibfield  {author} {\bibinfo {author} {\bibfnamefont {M.}~\bibnamefont
  {Rauscher}}, \bibinfo {author} {\bibfnamefont {A.}~\bibnamefont {Dominguez}},
  \bibinfo {author} {\bibfnamefont {M.}~\bibnamefont {Kr\"uger}}, \ and\
  \bibinfo {author} {\bibfnamefont {F.}~\bibnamefont {Penna}},\ }\href@noop {}
  {\bibfield  {journal} {\bibinfo  {journal} {J. Chem. Phys.}\ }\textbf
  {\bibinfo {volume} {127}},\ \bibinfo {pages} {244906} (\bibinfo {year}
  {2007})}\BibitemShut {NoStop}%
\bibitem [{\citenamefont {Zimmermann}\ \emph {et~al.}(2016)\citenamefont
  {Zimmermann}, \citenamefont {Smallenburg},\ and\ \citenamefont
  {L{\"o}wen}}]{Zimmermann16}%
  \BibitemOpen
  \bibfield  {author} {\bibinfo {author} {\bibfnamefont {U.}~\bibnamefont
  {Zimmermann}}, \bibinfo {author} {\bibfnamefont {F.}~\bibnamefont
  {Smallenburg}}, \ and\ \bibinfo {author} {\bibfnamefont {H.}~\bibnamefont
  {L{\"o}wen}},\ }\href@noop {} {\bibfield  {journal} {\bibinfo  {journal}
  {Journal of Physics: Condensed Matter}\ }\textbf {\bibinfo {volume} {28}},\
  \bibinfo {pages} {244019} (\bibinfo {year} {2016})}\BibitemShut {NoStop}%
\bibitem [{\citenamefont {H\"artel}\ \emph {et~al.}(2010)\citenamefont
  {H\"artel}, \citenamefont {Blaak},\ and\ \citenamefont {L\"owen}}]{Hartel10}%
  \BibitemOpen
  \bibfield  {author} {\bibinfo {author} {\bibfnamefont {A.}~\bibnamefont
  {H\"artel}}, \bibinfo {author} {\bibfnamefont {R.}~\bibnamefont {Blaak}}, \
  and\ \bibinfo {author} {\bibfnamefont {H.}~\bibnamefont {L\"owen}},\
  }\href@noop {} {\bibfield  {journal} {\bibinfo  {journal} {Phys. Rev. E}\
  }\textbf {\bibinfo {volume} {81}} (\bibinfo {year} {2010})}\BibitemShut
  {NoStop}%
\bibitem [{\citenamefont {Brader}\ and\ \citenamefont
  {Kr\"uger}(2011)}]{Brader_Kruger_2011}%
  \BibitemOpen
  \bibfield  {author} {\bibinfo {author} {\bibfnamefont {J.}~\bibnamefont
  {Brader}}\ and\ \bibinfo {author} {\bibfnamefont {M.}~\bibnamefont
  {Kr\"uger}},\ }\href@noop {} {\bibfield  {journal} {\bibinfo  {journal} {Mol.
  Phys.}\ }\textbf {\bibinfo {volume} {109}},\ \bibinfo {pages} {1029}
  (\bibinfo {year} {2011})}\BibitemShut {NoStop}%
\bibitem [{\citenamefont {Kr\"uger}\ and\ \citenamefont
  {Brader}(2011)}]{Brader_Kruger_2011_2}%
  \BibitemOpen
  \bibfield  {author} {\bibinfo {author} {\bibfnamefont {M.}~\bibnamefont
  {Kr\"uger}}\ and\ \bibinfo {author} {\bibfnamefont {J.}~\bibnamefont
  {Brader}},\ }\href@noop {} {\bibfield  {journal} {\bibinfo  {journal}
  {Eur.Phys.Lett}\ }\textbf {\bibinfo {volume} {96}},\ \bibinfo {pages} {68066}
  (\bibinfo {year} {2011})}\BibitemShut {NoStop}%
\bibitem [{\citenamefont {J.~Reinhardt}\ and\ \citenamefont
  {Brader}(2013)}]{Johannes_Joe_2013}%
  \BibitemOpen
  \bibfield  {author} {\bibinfo {author} {\bibfnamefont {F.~W.}\ \bibnamefont
  {J.~Reinhardt}}\ and\ \bibinfo {author} {\bibfnamefont {J.}~\bibnamefont
  {Brader}},\ }\href@noop {} {\bibfield  {journal} {\bibinfo  {journal}
  {Eur.Phys.Lett}\ }\textbf {\bibinfo {volume} {102}},\ \bibinfo {pages}
  {28011} (\bibinfo {year} {2013})}\BibitemShut {NoStop}%
\bibitem [{\citenamefont {Aerov}\ and\ \citenamefont
  {Kr\"uger}(2014)}]{AerovKrugerJCP2014}%
  \BibitemOpen
  \bibfield  {author} {\bibinfo {author} {\bibfnamefont {A.~A.}\ \bibnamefont
  {Aerov}}\ and\ \bibinfo {author} {\bibfnamefont {M.}~\bibnamefont
  {Kr\"uger}},\ }\href@noop {} {\bibfield  {journal} {\bibinfo  {journal} {J.
  Phys. Chem.}\ }\textbf {\bibinfo {volume} {140}},\ \bibinfo {pages} {094701}
  (\bibinfo {year} {2014})}\BibitemShut {NoStop}%
\bibitem [{\citenamefont {Aerov}\ and\ \citenamefont
  {Kr{\"u}ger}(2015)}]{aerov2015}%
  \BibitemOpen
  \bibfield  {author} {\bibinfo {author} {\bibfnamefont {A.~A.}\ \bibnamefont
  {Aerov}}\ and\ \bibinfo {author} {\bibfnamefont {M.}~\bibnamefont
  {Kr{\"u}ger}},\ }\href@noop {} {\bibfield  {journal} {\bibinfo  {journal}
  {Physical Review E}\ }\textbf {\bibinfo {volume} {92}},\ \bibinfo {pages}
  {042301} (\bibinfo {year} {2015})}\BibitemShut {NoStop}%
\bibitem [{\citenamefont {Menzel}\ \emph {et~al.}(2016)\citenamefont {Menzel},
  \citenamefont {Saha}, \citenamefont {Hoell},\ and\ \citenamefont
  {L{\"o}wen}}]{Menzel16}%
  \BibitemOpen
  \bibfield  {author} {\bibinfo {author} {\bibfnamefont {A.~M.}\ \bibnamefont
  {Menzel}}, \bibinfo {author} {\bibfnamefont {A.}~\bibnamefont {Saha}},
  \bibinfo {author} {\bibfnamefont {C.}~\bibnamefont {Hoell}}, \ and\ \bibinfo
  {author} {\bibfnamefont {H.}~\bibnamefont {L{\"o}wen}},\ }\href@noop {}
  {\bibfield  {journal} {\bibinfo  {journal} {The Journal of Chemical Physics}\
  }\textbf {\bibinfo {volume} {144}},\ \bibinfo {pages} {024115} (\bibinfo
  {year} {2016})}\BibitemShut {NoStop}%
\bibitem [{\citenamefont {Schmidt}\ and\ \citenamefont
  {Brader}(2013)}]{SchmidtBraderJCP_2013_power_func}%
  \BibitemOpen
  \bibfield  {author} {\bibinfo {author} {\bibfnamefont {M.}~\bibnamefont
  {Schmidt}}\ and\ \bibinfo {author} {\bibfnamefont {J.~M.}\ \bibnamefont
  {Brader}},\ }\href@noop {} {\bibfield  {journal} {\bibinfo  {journal} {J.
  Chem. Phys.}\ }\textbf {\bibinfo {volume} {138}},\ \bibinfo {pages} {214101}
  (\bibinfo {year} {2013})}\BibitemShut {NoStop}%
\bibitem [{\citenamefont {Brader}\ and\ \citenamefont
  {Schmidt}(2013)}]{Brader13}%
  \BibitemOpen
  \bibfield  {author} {\bibinfo {author} {\bibfnamefont {J.~M.}\ \bibnamefont
  {Brader}}\ and\ \bibinfo {author} {\bibfnamefont {M.}~\bibnamefont
  {Schmidt}},\ }\href@noop {} {\bibfield  {journal} {\bibinfo  {journal} {J.
  Chem. Phys.}\ }\textbf {\bibinfo {volume} {139}},\ \bibinfo {pages} {104108}
  (\bibinfo {year} {2013})}\BibitemShut {NoStop}%
\bibitem [{\citenamefont {Hopkins}\ \emph {et~al.}(2010)\citenamefont
  {Hopkins}, \citenamefont {Fortini}, \citenamefont {Archer},\ and\
  \citenamefont {Schmidt}}]{Hopkins10}%
  \BibitemOpen
  \bibfield  {author} {\bibinfo {author} {\bibfnamefont {P.}~\bibnamefont
  {Hopkins}}, \bibinfo {author} {\bibfnamefont {A.}~\bibnamefont {Fortini}},
  \bibinfo {author} {\bibfnamefont {A.~J.}\ \bibnamefont {Archer}}, \ and\
  \bibinfo {author} {\bibfnamefont {M.}~\bibnamefont {Schmidt}},\ }\href@noop
  {} {\bibfield  {journal} {\bibinfo  {journal} {The Journal of Chemical
  Physics}\ }\textbf {\bibinfo {volume} {133}} (\bibinfo {year}
  {2010})}\BibitemShut {NoStop}%
\bibitem [{\citenamefont {Archer}\ \emph {et~al.}(2007)\citenamefont {Archer},
  \citenamefont {Hopkins},\ and\ \citenamefont {Schmidt}}]{Archer10}%
  \BibitemOpen
  \bibfield  {author} {\bibinfo {author} {\bibfnamefont {A.~J.}\ \bibnamefont
  {Archer}}, \bibinfo {author} {\bibfnamefont {P.}~\bibnamefont {Hopkins}}, \
  and\ \bibinfo {author} {\bibfnamefont {M.}~\bibnamefont {Schmidt}},\ }\href
  {\doibase 10.1103/PhysRevE.75.040501} {\bibfield  {journal} {\bibinfo
  {journal} {Phys. Rev. E}\ }\textbf {\bibinfo {volume} {75}},\ \bibinfo
  {pages} {040501} (\bibinfo {year} {2007})}\BibitemShut {NoStop}%
\bibitem [{\citenamefont {Brader}\ and\ \citenamefont
  {Schmidt}(2015)}]{Brader15}%
  \BibitemOpen
  \bibfield  {author} {\bibinfo {author} {\bibfnamefont {J.}~\bibnamefont
  {Brader}}\ and\ \bibinfo {author} {\bibfnamefont {M.}~\bibnamefont
  {Schmidt}},\ }\href@noop {} {\bibfield  {journal} {\bibinfo  {journal} {J.
  Phys.: Condens. Matter}\ }\textbf {\bibinfo {volume} {27}},\ \bibinfo {pages}
  {194106} (\bibinfo {year} {2015})}\BibitemShut {NoStop}%
\bibitem [{\citenamefont {{W. van Megen}}\ and\ \citenamefont
  {Underwood}(1993)}]{Megen93}%
  \BibitemOpen
  \bibfield  {author} {\bibinfo {author} {\bibnamefont {{W. van Megen}}}\ and\
  \bibinfo {author} {\bibfnamefont {S.}~\bibnamefont {Underwood}},\ }\href@noop
  {} {\bibfield  {journal} {\bibinfo  {journal} {Phys. Rev. Lett.}\ }\textbf
  {\bibinfo {volume} {70}},\ \bibinfo {pages} {2766} (\bibinfo {year}
  {1993})}\BibitemShut {NoStop}%
\bibitem [{\citenamefont {Nyg\aa{}rd}\ \emph {et~al.}(2016)\citenamefont
  {Nyg\aa{}rd}, \citenamefont {Buitenhuis}, \citenamefont {Kagias},
  \citenamefont {Jefimovs}, \citenamefont {Zontone},\ and\ \citenamefont
  {Chushkin}}]{Nygard16}%
  \BibitemOpen
  \bibfield  {author} {\bibinfo {author} {\bibfnamefont {K.}~\bibnamefont
  {Nyg\aa{}rd}}, \bibinfo {author} {\bibfnamefont {J.}~\bibnamefont
  {Buitenhuis}}, \bibinfo {author} {\bibfnamefont {M.}~\bibnamefont {Kagias}},
  \bibinfo {author} {\bibfnamefont {K.}~\bibnamefont {Jefimovs}}, \bibinfo
  {author} {\bibfnamefont {F.}~\bibnamefont {Zontone}}, \ and\ \bibinfo
  {author} {\bibfnamefont {Y.}~\bibnamefont {Chushkin}},\ }\href {\doibase
  10.1103/PhysRevLett.116.167801} {\bibfield  {journal} {\bibinfo  {journal}
  {Phys. Rev. Lett.}\ }\textbf {\bibinfo {volume} {116}},\ \bibinfo {pages}
  {167801} (\bibinfo {year} {2016})}\BibitemShut {NoStop}%
\bibitem [{\citenamefont {Petravic}\ and\ \citenamefont
  {Harrowell}(2006)}]{Petravic2_2006_JCP_viscosity_lin_resp_between_planes}%
  \BibitemOpen
  \bibfield  {author} {\bibinfo {author} {\bibfnamefont {J.}~\bibnamefont
  {Petravic}}\ and\ \bibinfo {author} {\bibfnamefont {P.}~\bibnamefont
  {Harrowell}},\ }\href@noop {} {\bibfield  {journal} {\bibinfo  {journal} {J.
  Chem. Phys.}\ }\textbf {\bibinfo {volume} {124}},\ \bibinfo {pages} {044512}
  (\bibinfo {year} {2006})}\BibitemShut {NoStop}%
\bibitem [{\citenamefont {Klapp}\ \emph {et~al.}(2008)\citenamefont {Klapp},
  \citenamefont {Zeng}, \citenamefont {Qu},\ and\ \citenamefont
  {Klitzing}}]{Klapp2008PRL_SimCollFilm}%
  \BibitemOpen
  \bibfield  {author} {\bibinfo {author} {\bibfnamefont {S.~H.~L.}\
  \bibnamefont {Klapp}}, \bibinfo {author} {\bibfnamefont {Y.}~\bibnamefont
  {Zeng}}, \bibinfo {author} {\bibfnamefont {D.}~\bibnamefont {Qu}}, \ and\
  \bibinfo {author} {\bibfnamefont {R.}~\bibnamefont {Klitzing}},\ }\href@noop
  {} {\bibfield  {journal} {\bibinfo  {journal} {Phys. Rev. Lett.}\ }\textbf
  {\bibinfo {volume} {100}},\ \bibinfo {pages} {118303} (\bibinfo {year}
  {2008})}\BibitemShut {NoStop}%
\bibitem [{\citenamefont {Zhang}\ \emph {et~al.}(2004)\citenamefont {Zhang},
  \citenamefont {Todd},\ and\ \citenamefont {Travis}}]{Zhang2004SimPoisFl}%
  \BibitemOpen
  \bibfield  {author} {\bibinfo {author} {\bibfnamefont {J.}~\bibnamefont
  {Zhang}}, \bibinfo {author} {\bibfnamefont {B.~D.}\ \bibnamefont {Todd}}, \
  and\ \bibinfo {author} {\bibfnamefont {K.~P.}\ \bibnamefont {Travis}},\
  }\href@noop {} {\bibfield  {journal} {\bibinfo  {journal} {J. Chem. Phys.}\
  }\textbf {\bibinfo {volume} {121}},\ \bibinfo {pages} {10778} (\bibinfo
  {year} {2004})}\BibitemShut {NoStop}%
\bibitem [{\citenamefont {Ignatovich}\ and\ \citenamefont
  {Novotny}(2006)}]{IgnatovichPRL2006DetectingNanoPart}%
  \BibitemOpen
  \bibfield  {author} {\bibinfo {author} {\bibfnamefont {F.~V.}\ \bibnamefont
  {Ignatovich}}\ and\ \bibinfo {author} {\bibfnamefont {L.}~\bibnamefont
  {Novotny}},\ }\href@noop {} {\bibfield  {journal} {\bibinfo  {journal} {Phys.
  Rev. Lett.}\ }\textbf {\bibinfo {volume} {96}},\ \bibinfo {pages} {013901}
  (\bibinfo {year} {2006})}\BibitemShut {NoStop}%
\bibitem [{\citenamefont {Isa}\ \emph {et~al.}(2009)\citenamefont {Isa},
  \citenamefont {Besseling}, \citenamefont {Morozov},\ and\ \citenamefont
  {Poon}}]{Isa09}%
  \BibitemOpen
  \bibfield  {author} {\bibinfo {author} {\bibfnamefont {L.}~\bibnamefont
  {Isa}}, \bibinfo {author} {\bibfnamefont {R.}~\bibnamefont {Besseling}},
  \bibinfo {author} {\bibfnamefont {A.~N.}\ \bibnamefont {Morozov}}, \ and\
  \bibinfo {author} {\bibfnamefont {W.~C.~K.}\ \bibnamefont {Poon}},\
  }\href@noop {} {\bibfield  {journal} {\bibinfo  {journal} {Phys. Rev. Lett.}\
  }\textbf {\bibinfo {volume} {102}},\ \bibinfo {pages} {058302} (\bibinfo
  {year} {2009})}\BibitemShut {NoStop}%
\bibitem [{\citenamefont {Cheng}\ \emph {et~al.}(2011)\citenamefont {Cheng},
  \citenamefont {McCoy}, \citenamefont {Israelachvili},\ and\ \citenamefont
  {Cohen}}]{Cheng11}%
  \BibitemOpen
  \bibfield  {author} {\bibinfo {author} {\bibfnamefont {X.}~\bibnamefont
  {Cheng}}, \bibinfo {author} {\bibfnamefont {J.~H.}\ \bibnamefont {McCoy}},
  \bibinfo {author} {\bibfnamefont {J.~N.}\ \bibnamefont {Israelachvili}}, \
  and\ \bibinfo {author} {\bibfnamefont {I.}~\bibnamefont {Cohen}},\
  }\href@noop {} {\bibfield  {journal} {\bibinfo  {journal} {Science}\ }\textbf
  {\bibinfo {volume} {333}},\ \bibinfo {pages} {1276} (\bibinfo {year}
  {2011})}\BibitemShut {NoStop}%
\bibitem [{\citenamefont {Chevalier}\ \emph {et~al.}(2014)\citenamefont
  {Chevalier}, \citenamefont {Rodts}, \citenamefont {Chateau}, \citenamefont
  {Chevalier},\ and\ \citenamefont {Coussot}}]{Chevalier}%
  \BibitemOpen
  \bibfield  {author} {\bibinfo {author} {\bibfnamefont {T.}~\bibnamefont
  {Chevalier}}, \bibinfo {author} {\bibfnamefont {T.}~\bibnamefont {Rodts}},
  \bibinfo {author} {\bibfnamefont {X.}~\bibnamefont {Chateau}}, \bibinfo
  {author} {\bibfnamefont {C.}~\bibnamefont {Chevalier}}, \ and\ \bibinfo
  {author} {\bibfnamefont {P.}~\bibnamefont {Coussot}},\ }\href@noop {}
  {\bibfield  {journal} {\bibinfo  {journal} {Phys. Rev. E}\ }\textbf {\bibinfo
  {volume} {89}},\ \bibinfo {pages} {023002} (\bibinfo {year}
  {2014})}\BibitemShut {NoStop}%
\bibitem [{\citenamefont {Psaltis}\ \emph {et~al.}(2006)\citenamefont
  {Psaltis}, \citenamefont {Quake},\ and\ \citenamefont
  {Yang}}]{Psaltis2006_Nature_microfl_dev}%
  \BibitemOpen
  \bibfield  {author} {\bibinfo {author} {\bibfnamefont {D.}~\bibnamefont
  {Psaltis}}, \bibinfo {author} {\bibfnamefont {S.~R.}\ \bibnamefont {Quake}},
  \ and\ \bibinfo {author} {\bibfnamefont {C.}~\bibnamefont {Yang}},\
  }\href@noop {} {\bibfield  {journal} {\bibinfo  {journal} {Nature}\ }\textbf
  {\bibinfo {volume} {442}},\ \bibinfo {pages} {381} (\bibinfo {year}
  {2006})}\BibitemShut {NoStop}%
\bibitem [{\citenamefont {Sajeesh}\ and\ \citenamefont
  {Sen}(2014)}]{Sajeesh2014_rev_Part_sep_in_Mfl_dev}%
  \BibitemOpen
  \bibfield  {author} {\bibinfo {author} {\bibfnamefont {P.}~\bibnamefont
  {Sajeesh}}\ and\ \bibinfo {author} {\bibfnamefont {A.~K.}\ \bibnamefont
  {Sen}},\ }\href@noop {} {\bibfield  {journal} {\bibinfo  {journal}
  {Microfluid Nanofluid}\ }\textbf {\bibinfo {volume} {17}},\ \bibinfo {pages}
  {1} (\bibinfo {year} {2014})}\BibitemShut {NoStop}%
\bibitem [{\citenamefont {Li}\ \emph {et~al.}(2004)\citenamefont {Li},
  \citenamefont {Fang}, \citenamefont {Lin}, \citenamefont {Xu},\ and\
  \citenamefont {Chen}}]{Li_blood_fl_sim}%
  \BibitemOpen
  \bibfield  {author} {\bibinfo {author} {\bibfnamefont {H.}~\bibnamefont
  {Li}}, \bibinfo {author} {\bibfnamefont {H.}~\bibnamefont {Fang}}, \bibinfo
  {author} {\bibfnamefont {Z.}~\bibnamefont {Lin}}, \bibinfo {author}
  {\bibfnamefont {S.}~\bibnamefont {Xu}}, \ and\ \bibinfo {author}
  {\bibfnamefont {S.}~\bibnamefont {Chen}},\ }\href@noop {} {\bibfield
  {journal} {\bibinfo  {journal} {Phys. Rev. E}\ }\textbf {\bibinfo {volume}
  {69}},\ \bibinfo {pages} {031919} (\bibinfo {year} {2004})}\BibitemShut
  {NoStop}%
\bibitem [{\citenamefont {Zhou}\ and\ \citenamefont
  {Chang}(2005)}]{Zhou2005_capillary_blood_susp}%
  \BibitemOpen
  \bibfield  {author} {\bibinfo {author} {\bibfnamefont {J.}~\bibnamefont
  {Zhou}}\ and\ \bibinfo {author} {\bibfnamefont {H.~C.}\ \bibnamefont
  {Chang}},\ }\href@noop {} {\bibfield  {journal} {\bibinfo  {journal} {J.
  Colloid Interface Sci.}\ }\textbf {\bibinfo {volume} {287}},\ \bibinfo
  {pages} {647} (\bibinfo {year} {2005})}\BibitemShut {NoStop}%
\bibitem [{\citenamefont {Dean}(1996)}]{Dean96}%
  \BibitemOpen
  \bibfield  {author} {\bibinfo {author} {\bibfnamefont {D.~S.}\ \bibnamefont
  {Dean}},\ }\href {http://stacks.iop.org/0305-4470/29/i=24/a=001} {\bibfield
  {journal} {\bibinfo  {journal} {Journal of Physics A: Mathematical and
  General}\ }\textbf {\bibinfo {volume} {29}},\ \bibinfo {pages} {L613}
  (\bibinfo {year} {1996})}\BibitemShut {NoStop}%
\bibitem [{\citenamefont {Archer}\ and\ \citenamefont
  {Rauscher}(2004)}]{Archer04b}%
  \BibitemOpen
  \bibfield  {author} {\bibinfo {author} {\bibfnamefont {A.~J.}\ \bibnamefont
  {Archer}}\ and\ \bibinfo {author} {\bibfnamefont {M.}~\bibnamefont
  {Rauscher}},\ }\href@noop {} {\bibfield  {journal} {\bibinfo  {journal} {J.
  Phys. A: Math. Gen.}\ }\textbf {\bibinfo {volume} {37}},\ \bibinfo {pages}
  {9325} (\bibinfo {year} {2004})}\BibitemShut {NoStop}%
\bibitem [{\citenamefont {Donev}\ and\ \citenamefont
  {Vanden-Eijnden}(2014)}]{Donev14}%
  \BibitemOpen
  \bibfield  {author} {\bibinfo {author} {\bibfnamefont {A.}~\bibnamefont
  {Donev}}\ and\ \bibinfo {author} {\bibfnamefont {E.}~\bibnamefont
  {Vanden-Eijnden}},\ }\href@noop {} {\bibfield  {journal} {\bibinfo  {journal}
  {The Journal of Chemical Physics}\ }\textbf {\bibinfo {volume} {140}},\
  \bibinfo {pages} {234115} (\bibinfo {year} {2014})}\BibitemShut {NoStop}%
\bibitem [{\citenamefont {Bollinger}\ \emph {et~al.}(2015)\citenamefont
  {Bollinger}, \citenamefont {Jain},\ and\ \citenamefont
  {Truskett}}]{Bollinger15}%
  \BibitemOpen
  \bibfield  {author} {\bibinfo {author} {\bibfnamefont {J.~A.}\ \bibnamefont
  {Bollinger}}, \bibinfo {author} {\bibfnamefont {A.}~\bibnamefont {Jain}}, \
  and\ \bibinfo {author} {\bibfnamefont {T.~M.}\ \bibnamefont {Truskett}},\
  }\href@noop {} {\bibfield  {journal} {\bibinfo  {journal} {The Journal of
  Physical Chemistry B}\ }\textbf {\bibinfo {volume} {119}},\ \bibinfo {pages}
  {9103} (\bibinfo {year} {2015})}\BibitemShut {NoStop}%
\bibitem [{\citenamefont {de~las Heras}\ and\ \citenamefont
  {Schmidt}(2014)}]{Heras14}%
  \BibitemOpen
  \bibfield  {author} {\bibinfo {author} {\bibfnamefont {D.}~\bibnamefont
  {de~las Heras}}\ and\ \bibinfo {author} {\bibfnamefont {M.}~\bibnamefont
  {Schmidt}},\ }\href {\doibase 10.1103/PhysRevLett.113.238304} {\bibfield
  {journal} {\bibinfo  {journal} {Phys. Rev. Lett.}\ }\textbf {\bibinfo
  {volume} {113}},\ \bibinfo {pages} {238304} (\bibinfo {year}
  {2014})}\BibitemShut {NoStop}%
\bibitem [{\citenamefont {Kardar}(2007)}]{kardarbook}%
  \BibitemOpen
  \bibfield  {author} {\bibinfo {author} {\bibfnamefont {M.}~\bibnamefont
  {Kardar}},\ }\href@noop {} {\emph {\bibinfo {title} {Statistical physics of
  fields}}}\ (\bibinfo  {publisher} {Cambridge University Press},\ \bibinfo
  {year} {2007})\BibitemShut {NoStop}%
\bibitem [{\citenamefont {Onuki}(2002)}]{onukibook}%
  \BibitemOpen
  \bibfield  {author} {\bibinfo {author} {\bibfnamefont {A.}~\bibnamefont
  {Onuki}},\ }\href@noop {} {\emph {\bibinfo {title} {Phase transition
  dynamics}}}\ (\bibinfo  {publisher} {Cambridge University Press},\ \bibinfo
  {year} {2002})\BibitemShut {NoStop}%
\bibitem [{\citenamefont {Altland}\ and\ \citenamefont
  {Simons}(2010)}]{Altland}%
  \BibitemOpen
  \bibfield  {author} {\bibinfo {author} {\bibfnamefont {A.}~\bibnamefont
  {Altland}}\ and\ \bibinfo {author} {\bibfnamefont {B.}~\bibnamefont
  {Simons}},\ }\href {https://cds.cern.ch/record/1257155} {\emph {\bibinfo
  {title} {{Condensed matter field theory; 2nd ed.}}}}\ (\bibinfo  {publisher}
  {Cambridge Univ. Press},\ \bibinfo {address} {Cambridge},\ \bibinfo {year}
  {2010})\BibitemShut {NoStop}%
\bibitem [{\citenamefont {Hohenberg}\ and\ \citenamefont
  {Halperin}(1977)}]{hohenberg}%
  \BibitemOpen
  \bibfield  {author} {\bibinfo {author} {\bibfnamefont {P.}~\bibnamefont
  {Hohenberg}}\ and\ \bibinfo {author} {\bibfnamefont {B.}~\bibnamefont
  {Halperin}},\ }\href@noop {} {\bibfield  {journal} {\bibinfo  {journal} {Rev.
  Mod. Phys.}\ }\textbf {\bibinfo {volume} {49}} (\bibinfo {year}
  {1977})}\BibitemShut {NoStop}%
\bibitem [{Note1()}]{Note1}%
  \BibitemOpen
  \bibinfo {note} {We note that in our setup, the starting point, Eq.~\protect
  \textup {\hbox {\mathsurround \z@ \protect \normalfont (\ignorespaces \ref
  {sde}\unskip \@@italiccorr )}} is already a Langevin equation. Introducing
  the effective Hamiltonian thus gives rise to an additional reduction of
  explicit degrees of freedom.}\BibitemShut {Stop}%
\bibitem [{\citenamefont {Dean}\ and\ \citenamefont
  {Gopinathan}(2010)}]{deangopinathan2010PRE}%
  \BibitemOpen
  \bibfield  {author} {\bibinfo {author} {\bibfnamefont {D.~S.}\ \bibnamefont
  {Dean}}\ and\ \bibinfo {author} {\bibfnamefont {A.}~\bibnamefont
  {Gopinathan}},\ }\href@noop {} {\bibfield  {journal} {\bibinfo  {journal}
  {Phys. Rev. E}\ }\textbf {\bibinfo {volume} {81}},\ \bibinfo {pages} {041126}
  (\bibinfo {year} {2010})}\BibitemShut {NoStop}%
\bibitem [{\citenamefont {Velenich}\ \emph {et~al.}(2008)\citenamefont
  {Velenich}, \citenamefont {Chamon}, \citenamefont {Cugliandolo},\ and\
  \citenamefont {Kreimer}}]{velenich08}%
  \BibitemOpen
  \bibfield  {author} {\bibinfo {author} {\bibfnamefont {V.}~\bibnamefont
  {Velenich}}, \bibinfo {author} {\bibfnamefont {C.}~\bibnamefont {Chamon}},
  \bibinfo {author} {\bibfnamefont {L.~F.}\ \bibnamefont {Cugliandolo}}, \ and\
  \bibinfo {author} {\bibfnamefont {D.}~\bibnamefont {Kreimer}},\ }\href@noop
  {} {\bibfield  {journal} {\bibinfo  {journal} {J. Phys. A.}\ }\textbf
  {\bibinfo {volume} {41}},\ \bibinfo {pages} {235002} (\bibinfo {year}
  {2008})}\BibitemShut {NoStop}%
\bibitem [{\citenamefont {Fuchs}\ and\ \citenamefont {Cates}(2005)}]{fuchs}%
  \BibitemOpen
  \bibfield  {author} {\bibinfo {author} {\bibfnamefont {M.}~\bibnamefont
  {Fuchs}}\ and\ \bibinfo {author} {\bibfnamefont {M.~E.}\ \bibnamefont
  {Cates}},\ }\href@noop {} {\bibfield  {journal} {\bibinfo  {journal} {Journal
  of Physics: Condensed Matter}\ }\textbf {\bibinfo {volume} {17}},\ \bibinfo
  {pages} {S1681} (\bibinfo {year} {2005})}\BibitemShut {NoStop}%
\bibitem [{\citenamefont {Fuchs}\ and\ \citenamefont {Cates}(2009)}]{Fuchs09}%
  \BibitemOpen
  \bibfield  {author} {\bibinfo {author} {\bibfnamefont {M.}~\bibnamefont
  {Fuchs}}\ and\ \bibinfo {author} {\bibfnamefont {M.~E.}\ \bibnamefont
  {Cates}},\ }\href@noop {} {\bibfield  {journal} {\bibinfo  {journal} {J.
  Rheol.}\ }\textbf {\bibinfo {volume} {53}},\ \bibinfo {pages} {957} (\bibinfo
  {year} {2009})}\BibitemShut {NoStop}%
\bibitem [{\citenamefont {Cichocki}\ and\ \citenamefont
  {Felderhof}(1991)}]{Cichocki91}%
  \BibitemOpen
  \bibfield  {author} {\bibinfo {author} {\bibfnamefont {B.}~\bibnamefont
  {Cichocki}}\ and\ \bibinfo {author} {\bibfnamefont {B.~U.}\ \bibnamefont
  {Felderhof}},\ }\href@noop {} {\bibfield  {journal} {\bibinfo  {journal} {The
  Journal of Chemical Physics}\ }\textbf {\bibinfo {volume} {94}},\ \bibinfo
  {pages} {556} (\bibinfo {year} {1991})}\BibitemShut {NoStop}%
\bibitem [{\citenamefont {Cichocki}\ and\ \citenamefont
  {Felderhof}(1993)}]{Cichocki93}%
  \BibitemOpen
  \bibfield  {author} {\bibinfo {author} {\bibfnamefont {B.}~\bibnamefont
  {Cichocki}}\ and\ \bibinfo {author} {\bibfnamefont {B.~U.}\ \bibnamefont
  {Felderhof}},\ }\href@noop {} {\bibfield  {journal} {\bibinfo  {journal} {The
  Journal of Chemical Physics}\ }\textbf {\bibinfo {volume} {98}},\ \bibinfo
  {pages} {8186} (\bibinfo {year} {1993})}\BibitemShut {NoStop}%
\bibitem [{\citenamefont {Dean}\ and\ \citenamefont
  {Podgornik}(2014)}]{dean14}%
  \BibitemOpen
  \bibfield  {author} {\bibinfo {author} {\bibfnamefont {D.~S.}\ \bibnamefont
  {Dean}}\ and\ \bibinfo {author} {\bibfnamefont {R.}~\bibnamefont
  {Podgornik}},\ }\href@noop {} {\bibfield  {journal} {\bibinfo  {journal}
  {Phys. Rev. E}\ }\textbf {\bibinfo {volume} {89}},\ \bibinfo {pages} {032117}
  (\bibinfo {year} {2014})}\BibitemShut {NoStop}%
\bibitem [{\citenamefont {D\'emery}\ \emph {et~al.}(2014)\citenamefont
  {D\'emery}, \citenamefont {B\'enichou},\ and\ \citenamefont
  {Jacquin}}]{demery14}%
  \BibitemOpen
  \bibfield  {author} {\bibinfo {author} {\bibfnamefont {V.}~\bibnamefont
  {D\'emery}}, \bibinfo {author} {\bibfnamefont {O.}~\bibnamefont
  {B\'enichou}}, \ and\ \bibinfo {author} {\bibfnamefont {H.}~\bibnamefont
  {Jacquin}},\ }\href@noop {} {\bibfield  {journal} {\bibinfo  {journal} {New
  Journal of Physics}\ ,\ \bibinfo {pages} {053032}} (\bibinfo {year}
  {2014})}\BibitemShut {NoStop}%
\bibitem [{\citenamefont {Lu}\ \emph {et~al.}(2015)\citenamefont {Lu},
  \citenamefont {Dean},\ and\ \citenamefont {Podgornik}}]{lu15}%
  \BibitemOpen
  \bibfield  {author} {\bibinfo {author} {\bibfnamefont {B.-S.}\ \bibnamefont
  {Lu}}, \bibinfo {author} {\bibfnamefont {D.~S.}\ \bibnamefont {Dean}}, \ and\
  \bibinfo {author} {\bibfnamefont {R.}~\bibnamefont {Podgornik}},\ }\href@noop
  {} {\bibfield  {journal} {\bibinfo  {journal} {Europhys. Lett.}\ }\textbf
  {\bibinfo {volume} {112}},\ \bibinfo {pages} {20001} (\bibinfo {year}
  {2015})}\BibitemShut {NoStop}%
\bibitem [{\citenamefont {Dean}\ \emph {et~al.}(2016)\citenamefont {Dean},
  \citenamefont {Lu}, \citenamefont {Maggs},\ and\ \citenamefont
  {Podgornik}}]{dean16}%
  \BibitemOpen
  \bibfield  {author} {\bibinfo {author} {\bibfnamefont {D.~S.}\ \bibnamefont
  {Dean}}, \bibinfo {author} {\bibfnamefont {B.-S.}\ \bibnamefont {Lu}},
  \bibinfo {author} {\bibfnamefont {A.~C.}\ \bibnamefont {Maggs}}, \ and\
  \bibinfo {author} {\bibfnamefont {R.}~\bibnamefont {Podgornik}},\ }\href@noop
  {} {\bibfield  {journal} {\bibinfo  {journal} {Phys. Rev. Lett.}\ }\textbf
  {\bibinfo {volume} {116}},\ \bibinfo {pages} {240602} (\bibinfo {year}
  {2016})}\BibitemShut {NoStop}%
\bibitem [{\citenamefont {D\'emery}\ and\ \citenamefont
  {Dean}(2016)}]{demery16}%
  \BibitemOpen
  \bibfield  {author} {\bibinfo {author} {\bibfnamefont {V.}~\bibnamefont
  {D\'emery}}\ and\ \bibinfo {author} {\bibfnamefont {D.~S.}\ \bibnamefont
  {Dean}},\ }\href@noop {} {\bibfield  {journal} {\bibinfo  {journal} {J. Stat.
  Mech.}\ ,\ \bibinfo {pages} {023106}} (\bibinfo {year} {2016})}\BibitemShut
  {NoStop}%
\bibitem [{\citenamefont {Rohwer}\ \emph {et~al.}(2017)\citenamefont {Rohwer},
  \citenamefont {Kardar},\ and\ \citenamefont {Kr\"uger}}]{Rohwer17}%
  \BibitemOpen
  \bibfield  {author} {\bibinfo {author} {\bibfnamefont {C.~M.}\ \bibnamefont
  {Rohwer}}, \bibinfo {author} {\bibfnamefont {M.}~\bibnamefont {Kardar}}, \
  and\ \bibinfo {author} {\bibfnamefont {M.}~\bibnamefont {Kr\"uger}},\
  }\href@noop {} {\bibfield  {journal} {\bibinfo  {journal} {Phys. Rev. Lett.}\
  }\textbf {\bibinfo {volume} {118}},\ \bibinfo {pages} {015702} (\bibinfo
  {year} {2017})}\BibitemShut {NoStop}%
\end{thebibliography}

%
\end{document}